\documentclass[runningheads]{llncs}
\usepackage[T1]{fontenc}
\usepackage{subcaption}
\usepackage{graphicx}
\usepackage{amsmath,amssymb,amsfonts}
\usepackage{pifont}
\usepackage{booktabs}
\usepackage{multirow}
\usepackage[dvipsnames,svgnames,x11names,table]{xcolor}
\usepackage{hyperref}
\hypersetup{
    colorlinks=true,
    linkcolor=blue,
    filecolor=magenta,      
    urlcolor=cyan,
}
\usepackage[nameinlink]{cleveref}
\begin{document}
\title{Skin Lesion Segmentation Improved by Transformer-based Networks with Inter-scale Dependency Modeling}
\titlerunning{ISCF: Inter-scale Context Fusion}

\author{Sania Eskandari \and
	Janet Lumpp \and
    Luis {Sanchez Giraldo}
	}
\authorrunning{S. Eskandari et al.}

\institute{Department of Electrical Engineering, University of Kentucky, Lexington, USA \\
\email{ses235@uky.edu}}
\maketitle              % typeset the header of the contribution
% \blfootnote{$^\dag$ Corresponding Author.}

\begin{abstract}
Melanoma, a dangerous type of skin cancer resulting from abnormal skin cell growth, can be treated if detected early. Various approaches using Fully Convolutional Networks (FCNs) have been proposed, with the U-Net architecture being prominent To aid in its diagnosis through automatic skin lesion segmentation. However, the symmetrical U-Net model's reliance on convolutional operations hinders its ability to capture long-range dependencies crucial for accurate medical image segmentation. Several Transformer-based U-Net topologies have recently been created to overcome this limitation by replacing CNN blocks with different Transformer modules to capture local and global representations. Furthermore, the U-shaped structure is hampered by semantic gaps between the encoder and decoder. This study intends to increase the network's feature re-usability by carefully building the skip connection path. Integrating an already calculated attention affinity within the skip connection path improves the typical concatenation process utilized in the conventional skip connection path. As a result, we propose a U-shaped hierarchical Transformer-based structure for skin lesion segmentation and an Inter-scale Context Fusion (ISCF) method that uses attention correlations in each stage of the encoder to adaptively combine the contexts from each stage to mitigate semantic gaps. The findings from two skin lesion segmentation benchmarks support the ISCF module's applicability and effectiveness. The code is publicly available at \url{https://github.com/saniaesk/skin-lesion-segmentation}.

\end{abstract}
\section{Introduction}
The skin comprises three layers: the epidermis, dermis, and hypodermis~\cite{gordon2013skin}. When exposed to ultraviolet radiation from the sun, the epidermis produces melanin, which can be produced at an abnormal rate if too many melanocytes are present.  Malignant melanoma is a deadly form of skin cancer caused by the abnormal growth of melanocytes in the epidermis; in 2023, it was estimated that there would be 97,610 new cases of melanoma with a mortality rate of 8.18\%~\cite{siegel2023cancer}. The survival rate drops from 99\% to 25\% when melanoma is diagnosed at an advanced stage due to its aggressive nature~\cite{skincancer2023statistics,aghdam2022attention}. Therefore, early diagnosis is crucial in reducing the number of deaths from this disease. Dermoscopy has been introduced to improve the diagnosis procedure; it is a non-invasive technique that produces lighted and improved pictures of skin patches. Dermatologists use this method to detect skin cancer, which was formerly done by visual examination and manual screening, which was ineffective and time-consuming~\cite{yu2016automated}. Size, symmetry, boundary definition, and irregularity of lesion shape are essential for identifying skin cancer. Localization and delineation of lesions are required for both surgical excision and radiation treatment~\cite{skincancer2023statistics}. Manual delineation is a tedious and laborious task. Therefore, automatic segmentation becomes essential for developing pre and post-diagnosis processes in computer-aided diagnosis (CAD); however, automatic segmentation is challenging. %according to the \Cref{fig:skin_challenges}
Lighting and contrast difficulties, underlying inter-class similarities and intra-class variations, occlusions, artifacts, and various imaging tools impede automated skin lesion segmentation. The scarcity of large datasets with expert-generated ground-truth segmentation masks exacerbates the situation, hindering both model training and trustworthy assessment.

% \begin{figure}[!thb]
% 	\centering
% 	\includegraphics[width=0.6\textwidth]{./figures/skin_challenges.pdf}
% 	\caption{The visual challenges in dermoscopic images complicate the skin lesion segmentation task. The samples are selected from \textit{ISIC 2017} \cite{codella2018skin} dataset.}
% 	\label{fig:skin_challenges}
% \end{figure}

Before the Deep Learning (DL) era, most of the segmentation algorithms were based on hand-crafted and classical vision-based and conventional machine learning-based techniques. Celebi~et~al.~\cite{emre2013lesion} utilized adaptive thresholding, and in another study, \cite{emre2007unsupervised}, they investigated a region growing strategy, Erkol~et~al.~\cite{erkol2005automatic} applied the active contour method, and Hwang~et~al.~\cite{hwang2021segmentation} proposed a hybrid segmentation pipeline with an unsupervised clustering remedy where a hierarchical k-means with a level set technique was used. The aforementioned algorithms rely on human-engineered features, which may be challenging to construct and frequently have low invariance and discriminative ability.
% As a result, traditional segmentation algorithms may underperform on bigger and more complicated datasets. On the other hand, DL techniques provide a more effective answer to segmentation problems because they seamlessly integrate feature extraction with task-specific decision-making. 
As a result, the inadequacies of classic ML-based segmentation approaches emphasize the need for more advanced DL-based methods capable of handling complex data and producing more accurate results.

U-Net~\cite{ronneberger2015u} is the de facto in segmentation tasks. This DL-based framework is a cornerstone in medical image segmentation. U-Net is a hierarchical encoder-decoder framework comprising successive convolution operations in the encoding path, which downsample the spatial resolution while embedding the input space in a high dimensional space. The encoder provides a highly semantic representation, which is gradually upsampled in the decoding path to recover the input's spatial dimensions. Skip connections between the encoder and decoder in this design are used to mitigate the loss of spatial information, which is vital for segmentation tasks. Due to the modular design of the U-Net, thousands of variants of this network have been introduced that alleviate any shortcomings of the U-Net~\cite{azad2022medical}, \textit{e.g.,} U-Net++~\cite{zhou2018unet++}, H-DenseUNet~\cite{li2018h}, Attention U-Net~\cite{oktay2018attention}, \textit{etc}. U-Net++~\cite{zhou2018unet++} takes advantage of embedding nested U-Net structures in each stage by using a dense flow of semantic information from the encoder to the decoder with skip connections. Li et al. \cite{li2018h} replaced each naive convolutional encoder block with residual blocks besides using dense skip connections to extract more semantic information. However, applying successive convolution operations in dense structures still could not prevent the CNN-based U-shaped frameworks from suffering from having a limited receptive field. Attention U-Net~\cite{oktay2018attention} utilized the image-grid-based gating module that includes skip connections to let signals pass through and capture the gradient of relevant localization information from the encoder path before it merges with decoder features on the same scale. This strategy was the first seminal medical image segmentation study investigating the attention mechanism.

Transformer models in language translation tasks have been a huge success. This is related to its ability to calculate the self-affinity between the input tokens \cite{vaswani2017attention}. Dosovitskiy~et~al.~\cite{dosovitskiy2021an} proposed a Vision Transformer (ViT) which implements the attention mechanism on an image by partitioning the input images into a 1D sequence of patches to address the lack of globality of convolution operations. Soon, the Vision Transformer's ability to capture long-range dependencies in encoding the object's shape information inspired several studies that utilize the ViT for various tasks such as classification~\cite{dosovitskiy2021an} and segmentation~\cite{chen2021transunet}. However, due to the lack of intrinsic spatial inductive bias (impeding the ViT from capturing local representation) and the quadratic computational complexity of ViT (making ViT to be data hungry) with respect to the number of patches, the vanilla ViT performs poorly in dense prediction tasks like segmentation and object detection in comparison with the CNN models. Therefore, to mitigate the loss of local interactions within ViT, TransUNet~\cite{chen2021transunet}, TransBTS \cite{wang2021transbts}, UCTransNet~\cite{wang2022uctransnet}, and FAT-Net~\cite{wu2022fat} successfully bridge the CNN and Transformer designs in hybrid models. Due to the U-shaped design's success \cite{chen2021transunet,wang2021transbts,wu2022fat} hierarchical CNN-Transformer models tried capturing long and local dependencies simultaneously. The main drawbacks of these methods are that they still suffer from a high number of parameters and are dependent on the pre-trained weights to perform competitively. UCTransNet~\cite{wang2022uctransnet} investigated the semantic gap between encoder and decoder by designing a new Transformer-based module over skip connections to fuse the multi-scale spatial semantic information, but their method still requires pre-training weights.

Thus, various studies, such as the Efficient Transformer \cite{xie2021segformer} and the Swin Transformer \cite{liu2021swin}, have explored minimizing this computational burden to make ViTs suited for segmentation tasks by delving into the inner structure of the Transformer's multi-head self-attention (MHSA) calculation or by changing the tokenization process. Swin-Unet \cite{cao2023swin} is a hierarchical U-shaped pure Transformer structure that successfully utilized a Linear Swin Transformer as a main counterpart to segment the abdominal computer tomography inputs. Due to the shifting window strategy in Swin blocks, Swin-Unet captures the contextual information locally and is heavily dependent on pre-training weights. Huang~et~al.~\cite{huang2022missformer} applied the Efficient Transformer from~\cite{xie2021segformer} for medical image segmentation as a pure Transformer design, namely MISSFormer. Efficient Transformer~\cite{xie2021segformer} utilizes the irreversible downsampling step after the patch embedding to lessen the computational complexity, but this method suffers from loss of spatial information.

\begin{figure}[!thb]
	\centering
	\begin{subfigure}{0.67\textwidth}
		\centering
		\includegraphics[width=\textwidth]{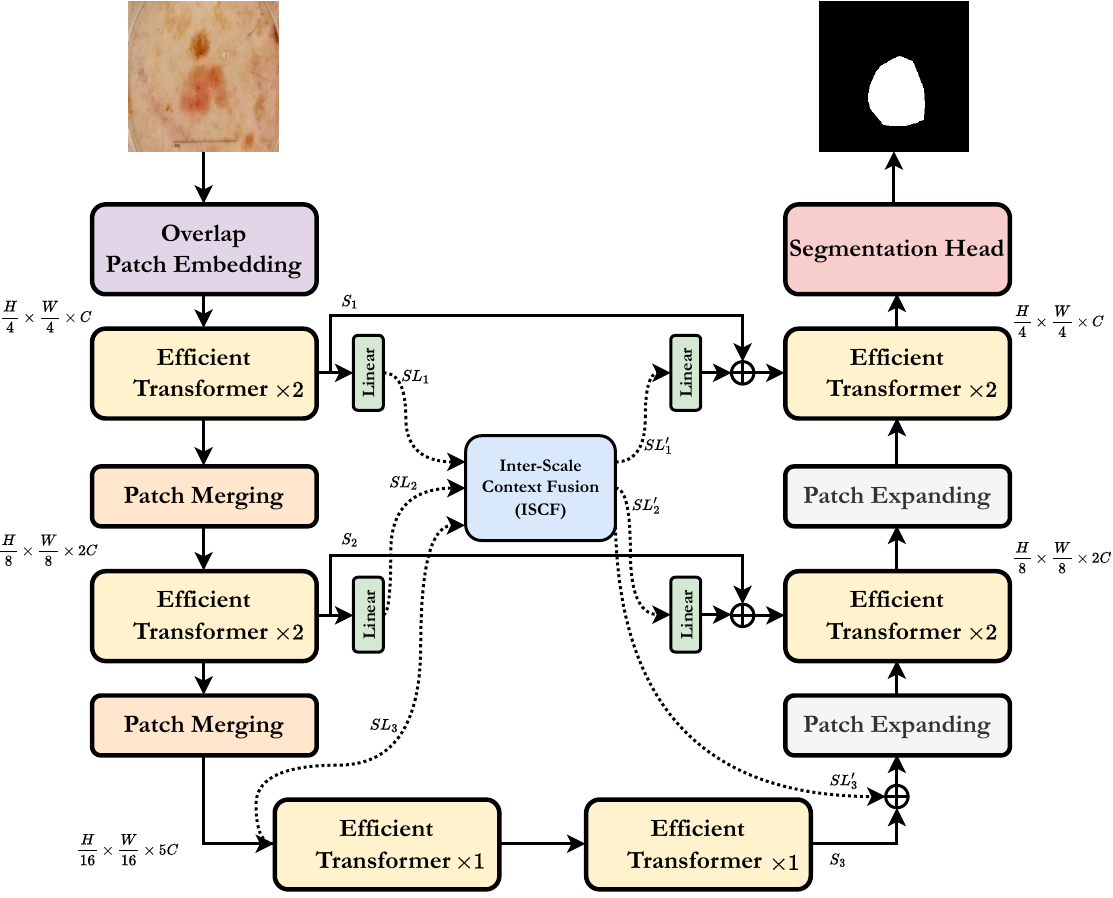}
		\caption{}
		\label{fig:method}
	\end{subfigure}
	\hfill
	\begin{subfigure}{0.27\textwidth}
		\centering
		\includegraphics[width=\textwidth]{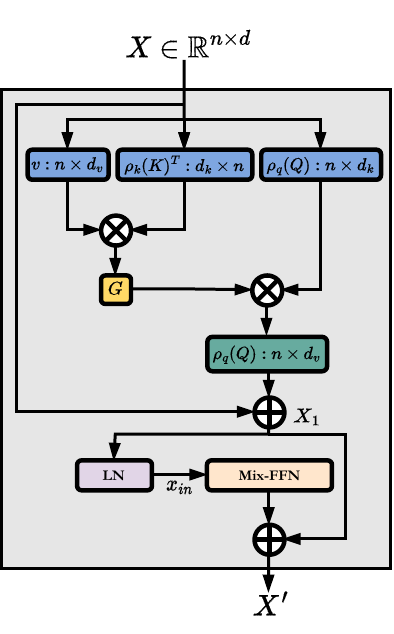}
		\caption{}
		\label{fig:efficientblock}
	\end{subfigure}
	\caption{(\textit{a}) Proposed method. Our U-shaped structure comprises modified Efficient attention. (\textit{b}) Our Efficient Transformer block.}
	\label{fig:method_and_efficientblock}
\end{figure}

\textbf{Our Contribution --}
To address the aforementioned deficiencies, we propose a new pure Transformer-based U-shaped structure that utilizes the efficient attention mechanism by Shen~et~al.~\cite{shen2021efficient} to capture the global context in linear complexity without redundant context extraction. Moreover, to shield the U-Net-like structures from the semantic gaps between the encoder and decoder, we devised a new module that uses an already calculated attention correlation at various scales to fuse the attention information for better localization. Our contributions are as follows: \ding{182} A novel Transformer-based structure in a U-shaped framework to capture the global dependency in an efficient manner without the need for pre-training weights (see \Cref{fig:method_and_efficientblock}). \ding{183} The design of a new skip connection module that integrates the multi-scale attention maps to lessen the encoder-decoder gap rather than the plain copy-and-paste skip connection paradigm, namely \textbf{I}nter-\textbf{S}cale \textbf{C}ontext \textbf{F}usion (\textbf{ISCF}). It is noteworthy that this study is the extended abstract version of \cite{eskandari2023interscale}. \ding{184} SOTA results on two public skin lesion segmentation datasets and publicly available implementation source code via \href{https://github.com/saniaesk/skin-lesion-segmentation}{GitHub}.

\section{Proposed Method}
Our overall framework proposed in \Cref{fig:method_and_efficientblock} is a convolution-free hierarchical U-shaped pure Transformer, designed for skin lesion segmentation. For an input image $\textrm{Im} \in \mathbb{R}^{H\times W \times C}$, where $H$, $W$, and $C$ denote the spatial dimensions and channels, respectively, our structure, as in \Cref{fig:method}, uses the patch merging and patch expanding strategies from \cite{liu2021swin,cao2023swin}. The patch embedding module extracts overlapping patch tokens of size $4 \times 4$ from the embedded tokens ($X \in \mathbb{R}^{n \times d}$) and then passes them through an encoder module with three stacked encoder blocks. Each block has two consecutive Efficient Transformer blocks and a patch merging layer that reduces spatial dimension (by merging $2 \times 2$ tokens) while doubling the channel dimension. The decoder expands the tokens by a factor of two in each block and integrates the output of each patch-expanding layer with the features forwarded by the fusion of the skip connection from the parallel encoder layer using ISCF. This approach enables the network to obtain a hierarchical representation.

\subsection{Efficient Attention}
Let $\mathbf{Q}$, $\mathbf{K}$, and $\mathbf{V}$  denote the Query, Key, and Value matrices that are produced by the embedded tokens $X$ in each stage, and $d$ is the embedding dimension. The standard self-attention equation is given by:
\begin{align}
	S(\mathbf{Q}, \mathbf{K}, \mathbf{V}) =\operatorname{softmax}\left(\frac{\mathbf{Q K^T}}{\sqrt{d}}\right) \mathbf{V}.\label{eq:self_attention}
\end{align}

The standard self-attention mechanism has a quadratic computational complexity ($\mathcal{O}(n^2)$), which limits its applicability in high-resolution tasks. Shen~et~al. \cite{shen2021efficient} proposed an approach called ``Efficient Attention'' that takes advantage of the fact that regular self-attention creates repetitive context matrix entries. They suggested a more efficient method for computing self-attention, as follows: %\Cref{eq:efficient-attention}: COULD NOT COMMENT THIS OUT WITH THE \DELETED MACRO!
\begin{align}
	\label{eq:efficient-attention}
	E(\mathbf{Q},\mathbf{K},\mathbf{V}) =  \mathbf{\rho_{q}}(\mathbf{Q})(\mathbf{\rho_{k}}(\mathbf{K})^{\mathbf{T}}\mathbf{V}),
\end{align}
where $\rho_{q}$ and $\rho_{k}$ are \textbf{Softmax}, normalization functions for the Queries and Keys. Shifting the order of multiplication drastically decreases the computation complexity to $\mathcal{O}(d^{2}n)$ when $d_{v} = d, d_{k} = \frac{d}{2}$, which is a typical setting (see \Cref{fig:efficientblock}). In contrast to naive dot-product attention (self-attention), efficient attention does not first compute pairwise similarities between points. Instead, the keys are represented as $d_{k}$ attention maps $\mathbf{k^{T}}_{j}$, with $j$ referring to position $j$ in the input feature. To sum up, according to \Cref{fig:efficientblock}, our Efficient Transformer block applies the following operations:
\begin{align}
	X_1 &= E(\mathbf{Q},\mathbf{K},\mathbf{V}) + X \\
	X^\prime &= \operatorname{Mix-FFN}(\operatorname{LN}(X_1)) + X_1,
\end{align}
where $\operatorname{LN}$ is a LayerNorm operation, and $\operatorname{Mix-FFN}$ is an enhanced Feed Forward Network (FFN) operation. It has been proven such a design can align features and make discriminative representations~\cite{huang2022missformer,xie2021segformer}, that mixes a $3 \times 3$ convolution and an MLP into each FFN. This operation actually plays as the dynamic positional encoding for each stage's Efficient transformer. 
% The detail about this module can be found in Appendix \ref{app:mix_ffn}.

\begin{figure}[!thb]
	\centering
	\includegraphics[width=0.8\textwidth]{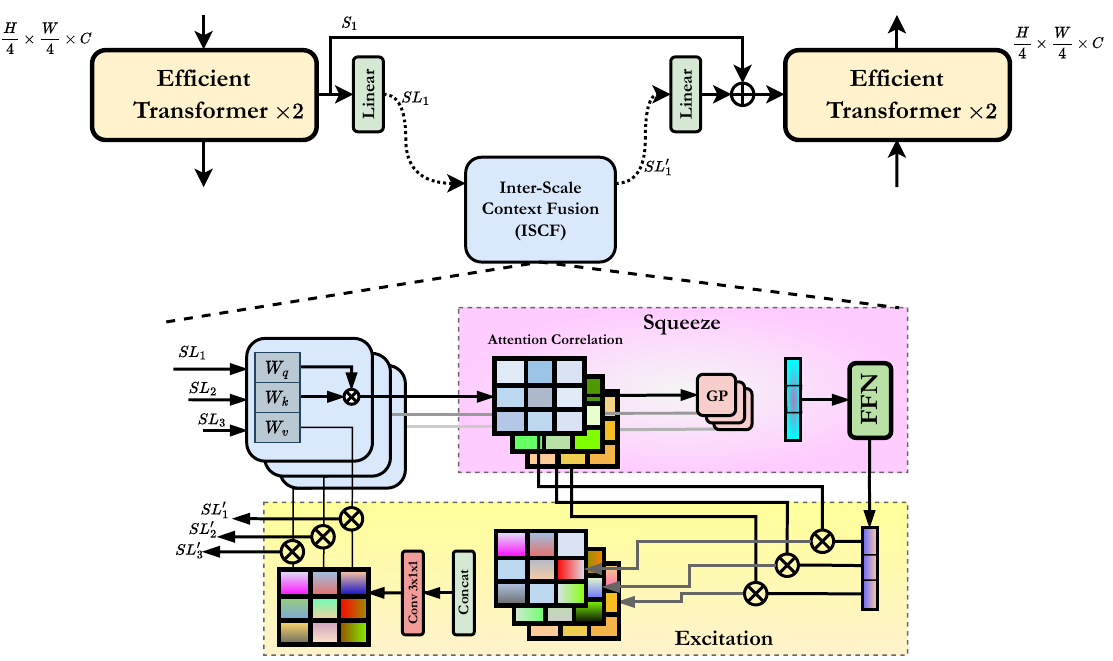}
	\caption{An overview of the Inter-Scale Context Fusion (ISCF) module. This module applies to already calculated attention maps in each stage's Transformer block and produces a general context over stages to compensate for the inter and intra-semantic gaps within the U-shaped design to improve the performance.}
	\label{fig:iscf}
 \vspace{-1em}
\end{figure}

\subsection{Inter-scale Context Fusion}
\vspace{-0.8em}
The ISCF module is displayed in \Cref{fig:iscf}. Instead of simply concatenating the features from the encoder and decoder layers, we devised a context fusion module to decrease the encoder-decoder semantic gap. Our proposed module not only can effectively provide spatial information to each decoder to recover fine-grained details when producing output masks, but also it does not require additional parameters for the model. Our proposed U-shaped structure is defined as a three stage multi-scale representation coupled with an ISCF module. Due to the hierarchical design of the structure, the attention maps' shape at each level differs from the next one. Therefore, we used a Linear layer in the first two stages two make the attention map sizes the same as the last stage. This operation is done at the output of the ISCF module to remap the attention maps to their original sizes. In the ISCF module, we utilize the Global Pooling (GP) operation to produce a single value for each stage's attention correlation and concatenate them, followed by an FFN to amalgamate the contribution of each global value into a scaling factor. Then each attention map applies the point-wise production with the corresponding scaling value and concatenates the resulting attention maps. Further, to adaptively fuse these global contexts to lessen the mentioned semantic gaps, a $3 \times 1 \times 1$ convolution is used. Finally, the resulting context fusion tensor adds to each plain skip connection from the encoder to the decoder to highlight the spatial localization for better segmentation results.

\section{Experimental Setup}
The PyTorch library was used to implement our proposed architecture and was run on a single RTX 3090 GPU. A batch size of 24 and an Adam solver with an empirically chosen learning rate of $1e-4$ were used for a $100$ epochs training. The loss function for the segmentation task was binary cross-entropy.

\subsection{Datasets}
\textbf{ISIC 2017 --} 
% Since 2016, International Skin Imaging Collaboration (ISIC) has released standard archive subsets as part of its \textit{Skin Lesion Analysis Towards Melanoma Detection Challenge}. The challenges in 2016, 2017, and 2018 included segmentation, feature extraction, and classification tasks, but those in 2019 and 2020 included classification. 
The ISIC 2017 dataset \cite{codella2018skin} comprises 2,000 skin dermoscopic images (cancer-positive and negative samples) with their corresponding annotations. We used 600 samples for the test set, 150 samples for data validation, and 1,250 samples for training. Each sample has an original size of $576 \times 767$ pixels. Images that are larger than $224 \times 224$ pixels are resized using the same pre-processing as \cite{alom2018recurrent}.

\noindent\textbf{ISIC 2018 --} ISIC 2018~\cite{codella2019skin} provides independent datasets for the classification and segmentation tasks for the first time, with 2,594 training (20\% melanomas, 72\% nevi, and 8\% seborrheic keratoses). Like prior approaches~\cite{alom2018recurrent}, we used 1,815 samples for training, 259 for validation, and 520 for testing. We downsize each sample image to $224 \times 224$ pixels from its original size of $2016 \times 3024$ pixels.

\subsection{Quantitative and Qualitative Results}
\begin{figure}
	\centering
	\includegraphics[width=0.8\textwidth]{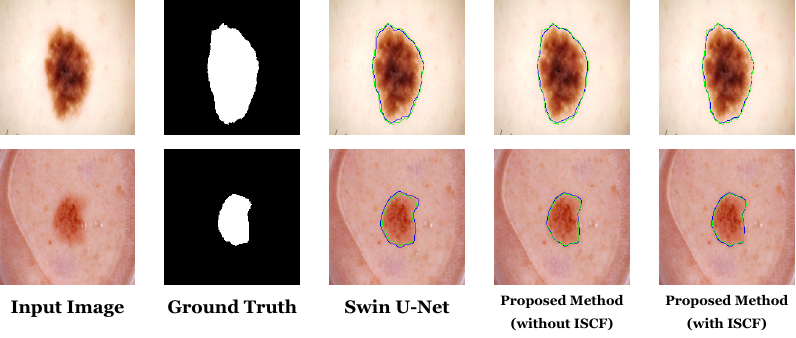}
	\caption{Qualitative results on \textit{ISIC 2018} dataset. \textcolor{green}{Green} represents the ground truth contour and \textcolor{blue}{blue} corresponds to the prediction mask contour.}
	\label{fig:qualitative}
\end{figure}

In \Cref{tab:results}, the quantitative results for our proposed method are displayed. We reported the performance of the model on the Dice score (DSC), sensitivity (SE), specificity (SP), and accuracy (ACC). Our results show that the proposed design can outperform SOTA methods without pre-training weights and having fewer parameters. In addition, \Cref{fig:qualitative} provides qualitative results that show the network performs well with respect to the ground truth segmentation and preserves the high-frequency details such as boundary information. It is evident that the boundary information with an Efficient Transformer and the ISCF module preserve high-frequency details effectively in comparison to the naive Swin~U-Net \cite{cao2023swin} and highlights the efficacy of the ISCF module in compensating the information gap between encoder and decoder. 

% \begin{figure}
% 	\centering
% 	\includegraphics[width=0.8\textwidth]{figures/skin_visualization.pdf}
% 	\caption{Qualitative results on \textit{ISIC 2018} dataset. \textcolor{green}{Green} represents the ground truth contour and \textcolor{blue}{blue} corresponds to the prediction mask contour.}
% 	\label{fig:qualitative}
% \end{figure}
\begin{table}[!thb]
	\vspace{-0.5cm}
	\caption{Performance comparison of the proposed method against SOTA approaches on the \textit{ISIC 2017}, and \textit{ISIC 2018} skin lesion segmentation tasks.} \label{tab:results}
	\centering
	\resizebox{\textwidth}{!}{
		\begin{tabular}{l || c || c c c c || c c c c}
			\hline
			\multirow{2}{*}{\textbf{Methods}} & \multirow{2}{*}{\textbf{ \# Params(M)}}& \multicolumn{4}{c||}{\textit{ISIC 2017}} & \multicolumn{4}{c}{\textit{ISIC 2018}} \\ \cline{3-6} \cline{7-10}
			&   & \textbf{DSC}   & \textbf{SE}    & \textbf{SP}   & \textbf{ACC} & \textbf{DSC}    & \textbf{SE}     & \textbf{SP}     & \textbf{ACC}    \\
			\hline
			U-Net \cite{ronneberger2015u}  &  14.8 & 0.8159 & 0.8172 & 0.9680 & 0.9164 & 0.8545 & 0.8800 & 0.9697  & 0.9404 \\
			Att U-Net \cite{oktay2018attention}  & 34.88 & 0.8082 & 0.7998 & 0.9776 & 0.9145 & 0.8566   & 0.8674  & \textbf{0.9863} & 0.9376  \\
			TransUNet \cite{chen2021transunet}  & 105.28 & 0.8123 & 0.8263 & 0.9577 & 0.9207 & 0.8499   & 0.8578   & 0.9653   & 0.9452    \\ 
			FAT-Net \cite{wu2022fat} &  28.75 & 0.8500 & 0.8392 & 0.9725 & 0.9326 & 0.8903  & 0.9100 & 0.9699   & 0.9578  \\
			Swin\,U-Net \cite{cao2023swin} & 82.3 & 0.9183 & 0.9142  & \textbf{0.9798} &  \textbf{0.9701} & 0.8946 & 0.9056  & 0.9798   & \textbf{0.9645}          \\
			\hline
			% Efficient Transformer  & 0.8904 & 0.8861 & 0.9698 & 0.9519  \\
			Efficient Transformer (without ISCF) & 22.31 & 0.8998 & 0.8834 & 0.9530 & 0.9578 & 0.8817& 0.8534&	0.9698& 0.9519 \\    
			% \rowcolor{cyan!10}
			\textbf{Efficient Transformer (with ISCF)} & \textbf{23.43} & \textbf{0.9257} & \textbf{0.9321} & 0.9793 & 0.9698 & \textbf{0.9136} & \textbf{0.9284} & 0.9723 & 0.9630     \\ \hline
		\end{tabular}
	}
	\vspace{-0.5cm}
\end{table}

\vspace{-1em}
\subsection{Ablation Study}
% \vspace{-0.8em}
 To investigate how various settings affect the performance of our model, we conducted ablation studies on the number of times our proposed attention strategy was used in skip connections, as well as on different input sizes. We examined the effects of incorporating our proposed attention module at 1/4, 1/8, and 1/16 resolution scales. We studied the impact on our model by selecting which pairs of scales $SL_i$ and $SL^\prime_i$ for $i \in \left\{ {1, 2 ,3} \right\}$. As shown in Table \ref{tab:ablationstudy}, the segmentation performance improved as we increased the number of skip connection modules (scales), supporting our attention module's efficacy in capturing a rich representation. We also evaluated the impact of increasing the input size to $384 \times 384$ and found that while it led to slightly improved segmentation results, it also incurred a higher computational cost.

\begin{table}[!thb]
	\centering
	\caption{Ablation study on the \textit{ISIC 2018} skin lesion segmentation benchmark.} \label{tab:ablationstudy}
	\resizebox{0.9\textwidth}{!}{
		\begin{tabular}{l c c c c} \toprule
			\textbf{Setting} &  \textbf{DSC} & \textbf{SE} & \textbf{SP} & \textbf{ACC} \\ \cmidrule{1-5}
			Using only $SL_i$ and $SL^\prime_i$ pairs ($i \in \{1\}$) & 0.9025 & 0.9305 & 0.9490 & 0.9451 \\
			Using $SL_i$ and $SL^\prime_i$ pairs ($i \in \{1, 2\}$)  & 0.9065 & 0.9274 & 0.9683 & 0.9597 \\
			Using $SL_i$ and $SL^\prime_i$ pairs ($i \in \{1, 2, 3\}$)  & 0.9136 & \textbf{0.9284} & 0.9723 & 0.9630 \\
			\cmidrule{1-5}
			Input image size~$384 \times 384$ & \textbf{0.9189} & 0.9265 & \textbf{0.9799} & \textbf{0.9659} \\
			\bottomrule
		\end{tabular}
	}
\end{table}

\section{Conclusion}
The semantic gap between the encoder and decoder in a U-shaped Transformer-based network can be mitigated by carefully recalibrating the already calculated attention maps from each stage. In this study, not only do we address the hierarchical semantic gap drawback, but also, our ISCF module highlights the importance of spatial attention. ISCF module is a plug-and-play and computation-friendly module that can effectively be applied to any Transformer-based architecture. The qualitative and quantitative results endorse the applicability of the proposed module.

\bibliographystyle{splncs04}
\bibliography{refs}

\begin{thebibliography}{10}
\providecommand{\url}[1]{\texttt{#1}}
\providecommand{\urlprefix}{URL }
\providecommand{\doi}[1]{https://doi.org/#1}

\bibitem{aghdam2022attention}
Aghdam, E.K., Azad, R., Zarvani, M., Merhof, D.: Attention swin u-net: Cross-contextual attention mechanism for skin lesion segmentation. arXiv preprint arXiv:2210.16898  (2022)

\bibitem{alom2018recurrent}
Alom, M.Z., Hasan, M., Yakopcic, C., Taha, T.M., Asari, V.K.: Recurrent residual convolutional neural network based on u-net (r2u-net) for medical image segmentation. arXiv preprint arXiv:1802.06955  (2018)

\bibitem{azad2022medical}
Azad, R., Aghdam, E.K., Rauland, A., Jia, Y., Avval, A.H., Bozorgpour, A., Karimijafarbigloo, S., Cohen, J.P., Adeli, E., Merhof, D.: Medical image segmentation review: The success of u-net. arXiv preprint arXiv:2211.14830  (2022)

\bibitem{cao2023swin}
Cao, H., Wang, Y., Chen, J., Jiang, D., Zhang, X., Tian, Q., Wang, M.: Swin-unet: Unet-like pure transformer for medical image segmentation. In: Computer Vision--ECCV 2022 Workshops: Tel Aviv, Israel, October 23--27, 2022, Proceedings, Part III. pp. 205--218. Springer (2023)

\bibitem{chen2021transunet}
Chen, J., Lu, Y., Yu, Q., Luo, X., Adeli, E., Wang, Y., Lu, L., Yuille, A.L., Zhou, Y.: Transunet: Transformers make strong encoders for medical image segmentation. arXiv preprint arXiv:2102.04306  (2021)

\bibitem{codella2019skin}
Codella, N., Rotemberg, V., Tschandl, P., Celebi, M.E., Dusza, S., Gutman, D., Helba, B., Kalloo, A., Liopyris, K., Marchetti, M., et~al.: Skin lesion analysis toward melanoma detection 2018: A challenge hosted by the international skin imaging collaboration (isic). arXiv preprint arXiv:1902.03368  (2019)

\bibitem{codella2018skin}
Codella, N.C., Gutman, D., Celebi, M.E., Helba, B., Marchetti, M.A., Dusza, S.W., Kalloo, A., Liopyris, K., Mishra, N., Kittler, H., et~al.: Skin lesion analysis toward melanoma detection: A challenge at the 2017 international symposium on biomedical imaging (isbi), hosted by the international skin imaging collaboration (isic). In: 2018 IEEE 15th international symposium on biomedical imaging (ISBI 2018). pp. 168--172. IEEE (2018)

\bibitem{dosovitskiy2021an}
Dosovitskiy, A., Beyer, L., Kolesnikov, A., Weissenborn, D., Zhai, X., Unterthiner, T., Dehghani, M., Minderer, M., Heigold, G., Gelly, S., Uszkoreit, J., Houlsby, N.: An image is worth 16x16 words: Transformers for image recognition at scale. In: International Conference on Learning Representations (2021), \url{https://openreview.net/forum?id=YicbFdNTTy}

\bibitem{emre2007unsupervised}
Emre~Celebi, M., Alp~Aslandogan, Y., Stoecker, W.V., Iyatomi, H., Oka, H., Chen, X.: Unsupervised border detection in dermoscopy images. Skin research and technology  \textbf{13}(4),  454--462 (2007)

\bibitem{emre2013lesion}
Emre~Celebi, M., Wen, Q., Hwang, S., Iyatomi, H., Schaefer, G.: Lesion border detection in dermoscopy images using ensembles of thresholding methods. Skin Research and Technology  \textbf{19}(1),  e252--e258 (2013)

\bibitem{erkol2005automatic}
Erkol, B., Moss, R.H., Joe~Stanley, R., Stoecker, W.V., Hvatum, E.: Automatic lesion boundary detection in dermoscopy images using gradient vector flow snakes. Skin Research and Technology  \textbf{11}(1),  17--26 (2005)

\bibitem{eskandari2023interscale}
Eskandari, S., Lumpp, J.: Inter-scale dependency modeling for skin lesion segmentation with transformer-based networks. In: Medical Imaging with Deep Learning, short paper track (2023), \url{https://openreview.net/forum?id=JExQEfV5um}

\bibitem{gordon2013skin}
Gordon, R.: Skin cancer: an overview of epidemiology and risk factors. In: Seminars in oncology nursing. vol.~29, pp. 160--169. Elsevier (2013)

\bibitem{huang2022missformer}
Huang, X., Deng, Z., Li, D., Yuan, X., Fu, Y.: Missformer: An effective transformer for 2d medical image segmentation. IEEE Transactions on Medical Imaging  (2022)

\bibitem{hwang2021segmentation}
Hwang, Y.N., Seo, M.J., Kim, S.M.: A segmentation of melanocytic skin lesions in dermoscopic and standard images using a hybrid two-stage approach. BioMed Research International  \textbf{2021},  1--19 (2021)

\bibitem{li2018h}
Li, X., Chen, H., Qi, X., Dou, Q., Fu, C.W., Heng, P.A.: H-denseunet: hybrid densely connected unet for liver and tumor segmentation from ct volumes. IEEE transactions on medical imaging  \textbf{37}(12),  2663--2674 (2018)

\bibitem{liu2021swin}
Liu, Z., Lin, Y., Cao, Y., Hu, H., Wei, Y., Zhang, Z., Lin, S., Guo, B.: Swin transformer: Hierarchical vision transformer using shifted windows. In: Proceedings of the IEEE/CVF international conference on computer vision. pp. 10012--10022 (2021)

\bibitem{oktay2018attention}
Oktay, O., Schlemper, J., Folgoc, L.L., Lee, M., Heinrich, M., Misawa, K., Mori, K., McDonagh, S., Hammerla, N.Y., Kainz, B., et~al.: Attention u-net: Learning where to look for the pancreas. arXiv preprint arXiv:1804.03999  (2018)

\bibitem{ronneberger2015u}
Ronneberger, O., Fischer, P., Brox, T.: U-net: Convolutional networks for biomedical image segmentation. In: International Conference on Medical image computing and computer-assisted intervention. pp. 234--241. Springer (2015)

\bibitem{shen2021efficient}
Shen, Z., Zhang, M., Zhao, H., Yi, S., Li, H.: Efficient attention: Attention with linear complexities. In: Proceedings of the IEEE/CVF winter conference on applications of computer vision. pp. 3531--3539 (2021)

\bibitem{siegel2023cancer}
Siegel, R.L., Miller, K.D., Wagle, N.S., Jemal, A.: Cancer statistics, 2023. CA: A Cancer Journal for Clinicians  \textbf{73}(1),  17--48 (2023). \doi{https://doi.org/10.3322/caac.21763}, \url{https://acsjournals.onlinelibrary.wiley.com/doi/abs/10.3322/caac.21763}

\bibitem{skincancer2023statistics}
Society, A.C.: Cancer facts and figures, 2023. \url{https://www.cancer.org/content/dam/cancer-org/research/cancer-facts-and-statistics/annual-cancer-facts-and-figures/2023/slideshow-2023-cancer-facts-and-figures.pptx} (2023), accessed: 2023-03-10

\bibitem{vaswani2017attention}
Vaswani, A., Shazeer, N., Parmar, N., Uszkoreit, J., Jones, L., Gomez, A.N., Kaiser, {\L}., Polosukhin, I.: Attention is all you need. Advances in neural information processing systems  \textbf{30} (2017)

\bibitem{wang2022uctransnet}
Wang, H., Cao, P., Wang, J., Zaiane, O.R.: Uctransnet: rethinking the skip connections in u-net from a channel-wise perspective with transformer. In: Proceedings of the AAAI conference on artificial intelligence. vol.~36, pp. 2441--2449 (2022)

\bibitem{wang2021transbts}
Wang, W., Chen, C., Ding, M., Yu, H., Zha, S., Li, J.: Transbts: Multimodal brain tumor segmentation using transformer. In: Medical Image Computing and Computer Assisted Intervention--MICCAI 2021: 24th International Conference, Strasbourg, France, September 27--October 1, 2021, Proceedings, Part I 24. pp. 109--119. Springer (2021)

\bibitem{wu2022fat}
Wu, H., Chen, S., Chen, G., Wang, W., Lei, B., Wen, Z.: Fat-net: Feature adaptive transformers for automated skin lesion segmentation. Medical Image Analysis  \textbf{76},  102327 (2022)

\bibitem{xie2021segformer}
Xie, E., Wang, W., Yu, Z., Anandkumar, A., Alvarez, J.M., Luo, P.: Segformer: Simple and efficient design for semantic segmentation with transformers. Advances in Neural Information Processing Systems  \textbf{34},  12077--12090 (2021)

\bibitem{yu2016automated}
Yu, L., Chen, H., Dou, Q., Qin, J., Heng, P.A.: Automated melanoma recognition in dermoscopy images via very deep residual networks. IEEE transactions on medical imaging  \textbf{36}(4),  994--1004 (2016)

\bibitem{zhou2018unet++}
Zhou, Z., Siddiquee, M.M.R., Tajbakhsh, N., Liang, J.: Unet++: A nested u-net architecture for medical image segmentation. In: Deep learning in medical image analysis and multimodal learning for clinical decision support, pp. 3--11. Springer (2018)

\end{thebibliography}

% \appendix
% \section{Norm \& MiX-FFN} \label{app:mix_ffn} 

% The resolution of Positional Encoding (PE) in ViT is fixed, and the hierarchical design of Transformer blocks suffers from this deficiency. Xie et al. \cite{xie2021segformer} proposed a Mix-FFN module to avoid the need for interpolating PEs, which proved in their study that it decreases the network's performance. Moreover, in Huang et al. \cite{huang2022missformer} comprehensive study, they endorsed the efficacy of this design, that such a design can align features and make discriminative representations. Therefore, we used the simple Mix-FFN in \cite{huang2022missformer} instead of a simple FFN block within the naive ViT \cite{dosovitskiy2021an} (see \Cref{fig:mix_ffn}). If $x_{in}$ is the output from the LayerNorm after the Efficient attention (see \Cref{fig:efficientblock}), then the calculation through \Cref{fig:mix_ffn} can be formulated as:

% % {\footnotesize
% 	\begin{align}
% 		x_{out} = \text{FC}(\text{GELU}(\text{LN}(\text{Depth-wise Conv}_{3\times 3}(\text{FC}(x_{in})) + \text{FC}(x_{in})))
% 	\end{align}
% 	% }
% \begin{figure}[!thb]
% 	\centering
% 	\includegraphics[width=0.13\textwidth]{./figures/mix_ffn.pdf}
% 	\caption{An overview of Mix-FFN module from the \cite{huang2022missformer}.}
% 	\label{fig:mix_ffn}
% \end{figure}

\end{document}